\documentclass[pra,twocolumn,showpacs]{revtex4}
\usepackage{bm,graphicx,amsmath}

\begin{document}

\title{Coupling internal atomic states in a two-component Bose-Einstein condensate via an optical lattice: Extended Mott-superfluid transitions}
\author{Jonas Larson}
\email{jolarson@kth.se}
\author{Jani-Petri Martikainen}
\affiliation{NORDITA, 106 91 Stockholm, Sweden}
\date{\today}

\begin{abstract}
An ultracold gas of coupled two-component atoms in an optical field
is studied. Due to the internal two-level structure of the atoms,
three competing energy terms exist; atomic kinetic, atomic internal,
and atom-atom interaction energies. A novel outcome of this
interplay, not present in the regular Bose-Hubbard model, is that in
the single band and tight binding approximations four different
phases appear: Two superfluid and two Mott phases. When passing
through the critical point between the two superfluid or the two
Mott phases, a swapping of the internal atomic populations takes
place. By means of the strong coupling expansion, we find the full
phase diagram for the four different phases.

\end{abstract}

 \maketitle

\section{Introduction}
Ultracold atoms in optical lattices provide a testbed of strongly
interacting many-body systems. The advantages of these systems
compared to corresponding models in condensed matter physics lie in
the high controllability in terms of purity, parameters, state
preparation, and state detection \cite{maciek}. Since the seminal
experiment by Bloch and co-workers, where the Mott-superfluid phase
transition was first realized \cite{bsf}, numerous experimental
achievements have been accomplished. For example, Anderson
localization of matter waves \cite{dis}, the Tonks gas characterized
by strong atom-atom interaction \cite{tonks}, and the Mott phase of
two fermionic compounds \cite{fermimott}.

Systems composed of atoms with internal level structure, such as
spinor gases, yield other interesting possibilities. It has been
demonstrated that the additional internal degree of freedom gives
rise to novel phases and quantum phase transitions \cite{spin1}.
Experiments on spinor condensates include, for example, coherent
transport in optical lattices \cite{cohtrans}, spin-mixing
\cite{spin2}, inherent spin tunneling \cite{spin3}, and symmetry
breaking \cite{spin4}. In these works, as for mixtures of atomic
species in optical lattices \cite{mix}, direct coupling between the
internal states is not considered. Coupling between the internal
atomic states may indeed render new phenomena. Krutitsky {\it et
al.} studied a $\Lambda$ configuration for the atoms, coupled by two
optical lattices \cite{multicomp1,multicomp2}. They particularly
showed that the Mott-superfluid phase transition may be of first
order and that ferromagnetic and antiferromagnetic types of
superfluid states can exist in such coupled models. Later in
Ref.~\cite{multicomp3}, Garc\`ia-Ripoll and co-workers considered
individual lattice configurations for internal (dressed) atomic
states. Coupling between atoms in these two lattices was induced by
atom collisions. In an earlier contribution \cite{jonasjani}, we
demonstrated an inherent topological phase transition in fermionic
systems originating from the interplay between internal and external
atomic degrees of freedom rather than kinetic and atom-atom
interaction energies as is normally the case for cold atoms in
optical lattices.

In this paper we examine a gas of ultracold interacting bosonic
$\Lambda$-atoms in an one dimensional optical lattice. The two
atomic transitions are driven by one laser field rendering the
optical lattice and another external laser lacking any spatial
dependence in its mode profile. The largely detuned excited atomic
level is adiabatically eliminated resulting in an effective coupled
$2\times2$ model. An essential point of the resulting Hamiltonian is
that decoupling of the two equations into two separate Schr\"odinger
(or Gross-Pitaevskii) equations is only possible in adiabatic or
diabatic regimes. The novel physics, however, is found in the
intermediate regime, which will be analyzed in this work. In the
next Section we show that the spectrum of the single particle
Hamiltonian possesses several interesting features, such as
anomalous dispersions with multiple local minima. The many-body
Hamiltonian is derived in Sec.~\ref{sec3} using an expansion of the
atom field operators in the lowest band Wannier functions. The
magnitudes of the Hamiltonian parameters, obtained from overlap
integrals containing the numerically computed Wannier functions,
allow us to collect the significant terms. In the parameter regimes
studied in this contribution, we end up with a Bose-Hubbard
Hamiltonian, where the boson operators create or annihilate atoms in
certain superpositions of their internal states. Utilizing the
strong coupling expansion, we are able to find the system phase
diagram in Sec.~\ref{ssec4b}.

In contrast to the usual tight-binding Bose-Hubbard model, our
system possesses four different phases: two superfluid states and
two Mott states characterized by different collective atomic
population inversions. Moreover, the two superfluid phases have
either zero or $\pi$ phase modulation between neighboring sites,
giving them a ferro- or antiferromagnetic property. It is
demonstrated that time-of-flight measurements would provide the
information needed to distinguish between the two superfluid states
as well as between the Mott states and the superfluids, while the
two Mott states can be identified via state dependent detection.

\section{Single particle Hamiltonian}\label{sec2}
In order to obtain a many-body theory, we first consider properties
of the corresponding single particle Hamiltonian. This enables us to
systematically derive the many-body counterpart which includes
atom-atom interaction in the next Section.

\begin{figure}[ht]
\begin{center}
\includegraphics[width=8cm]{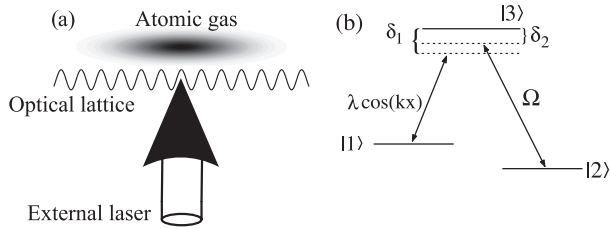}
\caption{Schematic configuration of system setup (a) and atom-laser
configuration (b). The optical lattice drives the
$|1\rangle\leftrightarrow|3\rangle$ atomic transition and the
external laser the $|2\rangle\leftrightarrow|3\rangle$ transition. }
\label{fig0}
\end{center}
\end{figure}

\subsection{The model system}
We consider an ultracold three-level $\Lambda$-atom with mass $m$
and internal levels $|i\rangle$, $i=1,2,3$, where $|1\rangle$ and
$|2\rangle$ are the two lower metastable states and $|3\rangle$ the
excited state. The atom moves in the presence of an one dimensional
optical lattice which couples the two atomic states 1 and 3 with an
effective coupling $\lambda$. The states 2 and 3 are coupled via an
``external" field with amplitude $\Omega$, which is furthermore
assumed constant over the extent of the atomic sample. Figure
\ref{fig0} details the system setup and the laser-atom configuration
we envision. Center-of-mass position and momentum are given by
$\hat{\tilde{x}}$ and $\hat{\tilde{p}}$ respectively. Both atomic
transitions are presumed highly detuned, with detunings $\delta_1$
and $\delta_2$ respectively, such that the excited state $|3\rangle$
can be adiabatically eliminated resulting in an effective two-level
model of the internal states $|1\rangle$ and $|2\rangle$. Following
standard procedures \cite{adel}, we derive the effective Hamiltonian
\begin{equation}\label{ham1}
\hat{H}_{sp}=\frac{\hat{\tilde{p}}^2}{2m}+\frac{\hbar\tilde{\Delta}}{2}\hat{\sigma}_z-\hbar\tilde{U}_1\cos(2k\hat{x})\hat{\sigma}_{11}+\hbar\tilde{U}\cos(k\hat{x})\hat{\sigma}_x,
\end{equation}
where
$\tilde{\Delta}=|\delta_1-\delta_2|-\Omega^2/\delta_2-\lambda^2/2\delta_1$
is an effective detuning taking into account for the constant Stark
shifts of states 1 and 2, $\tilde{U}_1=\lambda^2/2\delta_1$,
$\tilde{U}=\lambda\Omega(1/2\delta_1+1/2\delta_2)$, $k$ is the wave
number of the optical lattice, and for the $\sigma$-operators we
have $\hat{\sigma}_z=|2\rangle\langle2|-|1\rangle\langle1|$,
$\hat{\sigma}_x=|1\rangle\langle2|+|2\rangle\langle1|$ and
$\hat{\sigma}_{22}=|2\rangle\langle2|$. Note that the amplitudes
$\tilde{\Delta}$, $\tilde{U}_1$ and $\tilde{U}$ of the last three
terms of (\ref{ham1}) can be tuned independently within the validity
regime of the adiabatic elimination. For brevity, in the following
we will use dimensionless parameters. Letting $k^{-1}$ and
$E_r=\frac{\hbar^2k^2}{2m}$ set characteristic length and energy
scales, we scale the variables as
\begin{equation}
\hat{x}=k\hat{\tilde{x}},\hspace{0.6cm}\Delta=\frac{\hbar\tilde{\Delta}}{E_r},\hspace{0.6cm}U_1=\frac{\hbar\tilde{U}_1}{E_r},\hspace{0.6cm}U=\frac{\hbar\tilde{U}}{E_r}.
\end{equation}
In terms of scaled variables and in the $|1\rangle=\left[\begin{array}{c}0\\
1\end{array}\right]$ and $|2\rangle=\left[\begin{array}{c}1\\
0\end{array}\right]$ nomenclature, Eq.~(\ref{ham1}) becomes
\begin{equation}\label{ham2}
\hat{H}_{sp}=-\frac{\partial^2}{\partial x^2}+\left[\begin{array}{cc}
\displaystyle{\frac{\Delta}{2}} & U\cos(\hat{x}) \\
U\cos(\hat{x}) & -\displaystyle{\frac{\Delta}{2}}-U_1\cos(2\hat{x})\end{array}\right].
\end{equation}
It is worth emphasizing that with general parameters, this
Hamiltonian cannot be separated into two periodic Sch\"odinger
equations. There is no $x$-independent unitary matrix that would
diagonalize the $2\times2$ matrix of (\ref{ham2}). Thereby, the
corresponding matrix will not commute with the kinetic energy term,
causing non-diagonal couplings of the transformed Hamiltonian.
Indeed, this fact is a fundamental property for the results
presented in this work. This non-separability is different from most
earlier works on multi-component atoms in optical lattices, {\it
e.g.} in Ref.~\cite{multicomp3} a rotation decouples the system and
it is the atom-atom interaction that drives the coupling between two
effective equations, while in Ref.~\cite{multicomp1} the atoms
reside in a dark state.

The Hamiltonian (\ref{ham2}) is periodic with period
$\lambda=2\pi$ and thus, the operator $\hat{T}=\mathrm{e}^{\pm
i\lambda\hat{p}}$ is a constant of motion. Moreover, the
simultaneous inversion-displacement operator
\begin{equation}\label{sym}
\hat{I}=\hat{\sigma}_z\mathrm{e}^{\pm i\frac{\lambda}{2}\hat{p}}
\end{equation}
defines another symmetry of the Hamiltonian, where we explicitly
have $\hat{I}^2=\hat{T}$ \cite{jonaseff}. This additional invariant
reveals that the spectrum is most properly described within an
extended Brillouin zone with quasi momenta between -1 and 1. This
property has been discussed in greater detail in
Refs.~\cite{jonasjani,jonaseff}.

\subsection{Spectrum}
Labeling the momentum eigenstates by $|q\rangle$
($\hat{p}|q\rangle=q|q\rangle$), it is appropriate to divide the
{\it bare basis states} into two sets
\begin{equation}\label{basis}
\begin{array}{l}
|\varphi_\eta(q)\rangle=\left\{\begin{array}{l}
|q+\eta\rangle|1\rangle\hspace{1cm}\eta\,\,\mathrm{even}\\
|q+\eta\rangle|2\rangle\hspace{1cm}\eta\,\,\mathrm{odd}\end{array}\right.\\ \\
|\phi_\eta(q)\rangle=\left\{\begin{array}{l}
|q+\eta\rangle|2\rangle\hspace{1cm}\eta\,\,\mathrm{even}\\
|q+\eta\rangle|1\rangle\hspace{1cm}\eta\,\,\mathrm{odd},\end{array}\right.
\end{array}
\end{equation}
where $\eta$ is any integer and $q\in(-1,1]$. Due to the symmetry
defined by $\hat{I}$, the Hamiltonian (\ref{ham2}) is on
block-diagonal form within these states and consequently does not
couple basis states of different sets;
$\langle\varphi_{\eta'}(q')|\hat{H}|\phi_\eta(q)\rangle=0$. This
does not mean that the two-level structure of Eq.~(\ref{ham2}) has
been decoupled, but is merely a choice of basis states
(\ref{basis}). The analysis of Ref.~\cite{jonasjani} was carried out
by taking into account for the possibility of populating both sets
(\ref{basis}) simultaneously. Here we restrict ourself to one of the
blocks of the Hamiltonian, namely discard the states
$|\phi_\eta(q)\rangle$. Physically this implies that we assume a
particular initial state of the atom. More precisely, assuming the
atom to be ultracold with a momentum within the lowest Bloch band
and initial internal state $|1\rangle$. Such a constrain on the
initial atomic state seems reasonable within experimental
feasibility. We point out though, that by limiting our investigation
to a single set of (\ref{basis}), we do not overlook the physical
phenomena of interest for this work. One important observation is
that scattering between atoms may cause the two sets of basis states
$|\varphi_\eta(q)\rangle$ and $|\phi_\eta(q)\rangle$ to become
coupled, even if they are disconnected by the Hamiltonian. We will
return to this issue in Sec.~\ref{ssec4b}.

\begin{figure*}[ht]
\begin{center}
\includegraphics[width=14cm]{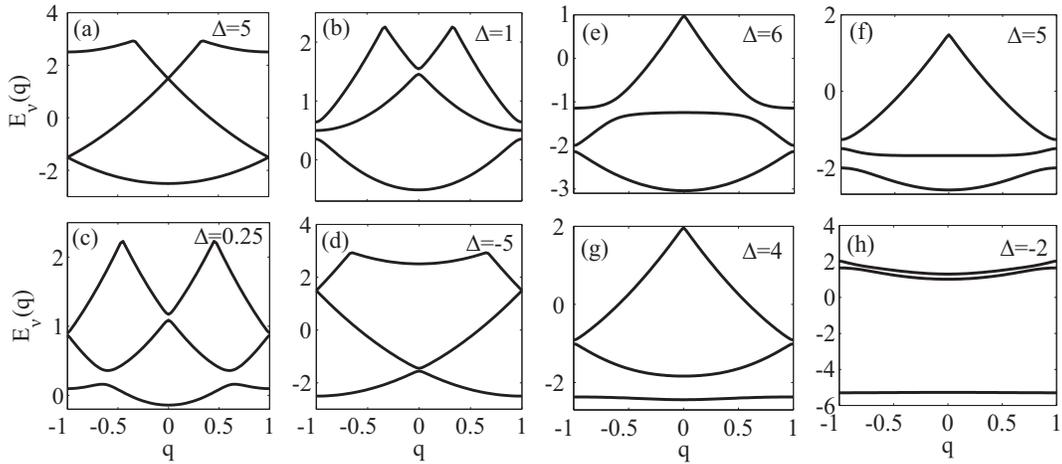}
\caption{Examples of the three lowest energy bands of Hamiltonian
(\ref{ham2}). In (a)-(d) we have $U>U_1$ ($U=0.05$ and $U_1=0.025$),
while in (e)-(h) $U<U_1$ ($U=0.1$ and $U_1=2$). The multiple number
of local minima of the lowest band is clear in (c). The large
detuning case is shown in (a), in which the lowest band has the
regular form. All parameters are dimensionless.} \label{fig1}
\end{center}
\end{figure*}

As a periodic problem, the eigenstates of $\hat{H}$ will be of {\it
Bloch form} imprinted with two quantum numbers, {\it band index}
$\nu=1,\,2,\,3,...$ and {\it quasi momentum} $q\in(-1,1]$,
\begin{equation}
\hat{H}|\psi_\nu(q)\rangle=E^\nu(q)|\psi_\nu(q)\rangle,
\end{equation}
where $E^\nu(q)$ is the $\nu$'th {\it Bloch band's} dispersion
curve. Due to the coupled two-level structure, the dispersions may
have anomalous shapes with multiple local minima
\cite{jonasjani,jonaseff}. This should be compared to regular energy
bands, by which we mean that either $dE^\nu(q)/dq\geq0$ or
$dE^\nu(q)/dq\leq0$ for $0\leq q\leq1$. Figure \ref{fig1} presents
several examples of the first three bands. The atypical forms of the
dispersions are clearly visible. In \cite{jonasjani} it was
demonstrated how such properties of the spectrum render a
topological PT for fermions and a first order PT for bosons.
Whenever $U\gg U_1$ (plots (a)-(d)), situations similar to the ones
presented in Ref.~\cite{jonasjani} are recovered. We remind that
here, however, we restrict the spectrum to only one subset in
Eq.~(\ref{basis}). As was found in \cite{jonasjani}, the lowest
dispersion curve can possess several local minima in this regime,
which is shown in Fig.~\ref{fig1} (c). Note that for $\Delta\gg0$,
the regular spectrum for the lowest band is recovered, while for
$\Delta\ll0$ the spectrum is ''shifted" by one unit of momentum
arising from the fact that we assumed the atoms to be initially in
their internal state $|1\rangle$. The reverse is obtained by
considering atoms initially in $|2\rangle$ instead. Observe further
that also the excited bands in the plots for large detuning possess
an irregular structure.

The situation is qualitatively different if instead $\Delta,\,U_1\gg
U$. Here, the internal states of the atoms are only weakly coupled
and we can approximate the spectrum by consisting of free particles
(atoms in internal state $|2\rangle$) and atoms in a potential
$U_1\cos(2x)$ (atoms in internal state $|1\rangle$). The
corresponding two spectrums may overlap forming unusual appearances.
For large $U_1$, the lowest dispersion curves, corresponding to
$|1\rangle$-atoms, are almost flat. Thus, slight non-zero repulsive
interaction between the atoms in such a band would cause an
insulating state. However, the size and sign of the detuning
$\Delta$ determines if the overall lowest band belongs to
$|1\rangle$ or $|2\rangle$ atoms and therefore if the state is in a
superfluid or a Mott state, provided repulsive atom-atom
interaction. These conclusions are verified in Fig. \ref{fig1}
(e)-(h), where in (e) the lowest dispersion curve is approximately
parabolic while in (h) it is almost flat. Similar structure of the
dispersions, mixture of narrow and wide energy bands, was also
encountered in honeycomb lattices \cite{hclattice}. We further note
that anomalous lowest band dispersions can also be achieved via
Bose-Hubbard models in square lattices beyond the tight binding
approximation~\cite{thomas}.

\section{The Bose-Hubbard Hamiltonian}\label{sec3}
One among many prototype models of many-body physics and the study of
quantum PTs is the {\it Bose-Hubbard model} \cite{fisher}. The
dynamics is driven by two terms representing hopping between sites
and on-site interaction between the particles. For ultracold bosonic
atoms in optical lattices, the analysis is most often restricted to
considering only the lowest band i.e. {\it single band approximation} and
to hopping only between neighboring sites i.e. {\it tight binding
approximation}. In this paper we impose these
approximations, and the validity of such assumptions will be
discussed in detail in Sec.~\ref{ssec4c}.

\subsection{Second quantization}\label{ssec3a}
The many-body Hamiltonian is given by
\begin{equation}\label{gp}
\begin{array}{lll}
\hat{H}\! & \!=\! & \!\displaystyle{\!\int\!\hat{\Psi}^\dagger(x)\!\cdot\!\Bigg\{\!-\frac{d^2}{dx^2}+\!\!\left[\begin{array}{cc}
\displaystyle{\frac{\Delta}{2}} & U\cos(\hat{x}) \\ \!\!
U\!\cos(\hat{x}) & -\displaystyle{\frac{\Delta}{2}}\!-U_1\!\cos(2\hat{x})\!\!\end{array}\right]}\\ \\
& & \displaystyle{+\frac{1}{2}\hat{\Psi}^\dagger(x)\cdot\mathbf{g}\cdot\hat{\Psi}(x)\Bigg\}\cdot\hat{\Psi}(x)dx},
\end{array}
\end{equation}
where $\hat{\Psi}(x)$ and $\hat{\Psi}^\dagger(x)$ are atomic spinor
annihilation and creation field operators respectively and
\begin{equation}
\mathbf{g}=\left[\begin{array}{cc} g_{11} & g_{12} \\
g_{12} & g_{22}\end{array}\right]
\end{equation}
is the scaled onsite interaction matrix with amplitudes $g_{ij}$. We
will make the approximation of assuming equal scattering amplitudes
between the internal condensate states $|1\rangle$ and $|2\rangle$,
$g_{11}=g_{22}=g$, and letting $g_{12}=0$. The atomic field operator
is conveniently expressed in terms of Wannier functions as
\begin{equation}\label{wanexp}
\hat{\Psi}(x)=\sum_i\mathrm{e}^{i\theta_i}\mathbf{w}_i(x)\hat{b}_{i},
\end{equation}
where $\mathbf{w}_i(x)=[w_1(x-x_i)\,\, w_2(x-x_i)]^T$ is the lowest
band's two-component Wannier function located at $x_i$ and $i$ runs
over all lattice sites, $\hat{b}_{i}$ is the bosonic annihilation
operator for the lowest band at site $i$ and the $\theta_i$'s are
phases that will be determined later. We remind that we consider an
one dimensional lattice so that the Wannier functions have a single
spatial coordinate. From our definition of the boson operators
$\hat{b}_i^\dagger$, it is clear that these have the meaning of
creating an atom with wave function
$\psi_{atom}(x-x_i)=w_1(x-x_i)|1\rangle+w_2(x-x_i)|2\rangle$. In
general, both the states $|1\rangle$ and $|2\rangle$ are populated
in such a state, and the populations depend highly on the system
parameters. Using (\ref{wanexp}), we derive the second quantized
Hamiltonian
\begin{equation}
\begin{array}{lll}
\hat{H}_{sb} & = & \displaystyle{-\sum_{i,j}J_{ij}\,\hat{b}_i^\dagger\hat{b}_j\mathrm{e}^{i(\theta_j-\theta_i)}}\\ \\
& & +\displaystyle{\frac{1}{2}\sum_{ijkl}G_{ijkl}\hat{b}_i^\dagger\hat{b}_j^\dagger\hat{b}_k\hat{b}_l-\mu\sum_i\hat{n}_i}.
\end{array}
\end{equation}
Here we have introduced the chemical potential $\mu$, and
\begin{equation}\label{overint}
\begin{array}{l}
\displaystyle{J_{ij}=-\int\mathbf{w}_i^\dagger(x)\cdot\hat{H}_{sp}\cdot\mathbf{w}_j(x)dx},\\ \\
\displaystyle{G_{ijkl}=g\int\mathbf{w}_i^\dagger(x)\cdot\left(\mathbf{w}_j^\dagger(x)\cdot\mathbf{w}_k(x)\right)\cdot\mathbf{w}_l(x)dx}
\end{array}
\end{equation}
are the overlap integrals determining the strength of hopping and
atom-atom interaction as function of the system parameters $\Delta$,
$U_1$ and $U$. So far, no tight binding approximation has been
applied. However, imposing the single band approximation normally
motivates the use of the tight binding approximation
\cite{jonasbcn}, in which case one obtains a Bose-Hubbard
Hamiltonian
\begin{equation}\label{eham}
\begin{array}{lll}
\hat{H}_{BH} & = & \displaystyle{-J_1\sum_{\langle i,j\rangle}\,\hat{b}_i^\dagger\hat{b}_j\mathrm{e}^{i\theta_{ji}}}\\ \\
& & +\displaystyle{\frac{G_0}{2}\sum_{i}\hat{n}_i(\hat{n}_i-1)-\mu\sum_i\hat{n}_i},
\end{array}
\end{equation}
where $\theta_{ji}=\theta_j-\theta_i$. The first sum runs over
nearest neighbors, the coefficients $J_1\equiv J_{ii+1}$ and
$G_0\equiv G_{iiii}$, and $\hat{n}_i=\hat{b}_i^\dagger\hat{b}_i$.
The phases $\theta_{ji}$ are chosen such that the total energy is
minimized, giving
\begin{equation}
\theta_{ji}=\left\{\begin{array}{ll}
0,\hspace{0.5cm} & J_1>0 \\
\pi,\hspace{0.5cm} & J_1<0.
\end{array}\right.
\end{equation}

The positions $x_i$ of the $i$'th Wannier function are not {\it a
priori} given in the present model. In general, the $x_i$'s are
taken to coincide with the minima of the effective potential. Here,
however, the coupled dynamics provide a situation where well defined
potentials cannot be ascribed single internal atomic states. These
issues were analyzed in more detail in \cite{jonasjani} and it was
in particular found that $x_i=n\pi$ for any integer $n$ renders
Wannier functions having familiar shapes, which for a deep lattice
approximate the harmonic oscillator eigenstates. This finding may be
motivated by the following argument. For positive detuning and
$\Delta\gg U\gg U_1$, the lowest dispersion has the regular form
(see Fig.~\ref{fig1} (a)), and since the initial atomic states are
chosen to be $|1\rangle$ the effective lattice potential is in the
adiabatic regime $U_{eff}(x)\propto\cos^2(x)$ which possesses its
minima for $x_i=n\pi$. Assuming a deep lattice, the Wannier
functions then attain, to a good accuracy, the forms of the
corresponding harmonic oscillator eigenfunctions in this limit.
Note, however, that our Wannier functions are still spinors, but in
this limit the $|1\rangle$ internal state is predominantly
populated. We have found that the Wannier functions preserve their
typical forms even for decreasing detunings $\Delta$ provided we
pick $x_i=n\pi$. By this we mean that both constituent Wannier
functions have an approximate Gaussian shape for relatively strong
lattices. This is indeed only true for $x_i=n\pi$ once we have
restricted the analysis to the basis set $|\varphi_\eta(q)\rangle$
of Eq.~(\ref{basis}). Any other choice of $x_i$ results in Wannier
functions with atypical shapes and being less localized. Therefore,
in the following we will choose $x_i=n\pi$.

\begin{figure}[ht]
\begin{center}
\includegraphics[width=8cm]{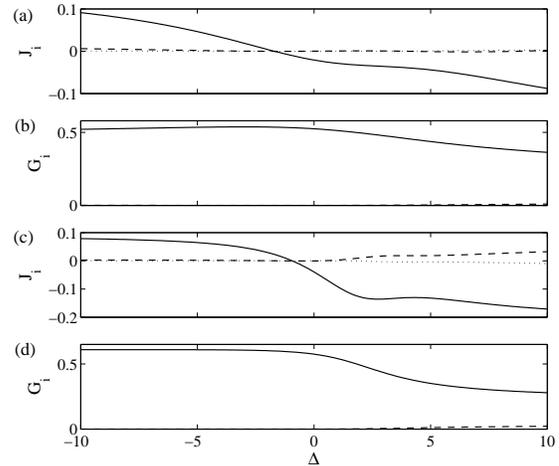}
\caption{Two examples of the first three hopping parameters $J_1$
(solid line), $J_2$ (dashed line) and $J_3$ (dotted line) as
function of $\Delta$ in (a) and (c), and $G_0$ (solid line) and
$G_1$ (dashed line) in (b) and (d). In both examples, $J_1$ changes
sign. The dimensionless parameters are $g=1$, $U=1$, and $U_1=0.5$
in (a) and (b), and $g=1$, $U=0.5$, and $U_1=1$ in (c) and (d).}
\label{fig2}
\end{center}
\end{figure}

The anomalous form of the lowest dispersion curve implies that the
nearest neighbor hopping parameter $J_1$ can attain both positive
and negative values \cite{multicomp1}. The coupling parameters,
given by the various overlap integrals (\ref{overint}), are
calculated using the spinor Wannier functions obtained numerically
from diagonalization of the Hamiltonian (\ref{ham2}). We display, as
a function of $\Delta$, $J_i$ ($i=1,2,3$) in Fig.~\ref{fig2} (a) and
(c), while (b) and (d) show $G_0$ and $G_1$, where $G_1=G_{ijji}$
and $j=i\pm1$. The coefficients $J_2\equiv J_{ii+2}$ and $J_3\equiv
J_{ii+3}$ describe the next and next-next nearest neighbor tunneling
strengths. The parameters are $U=1$ and $U_1=0.5$ in (a) and (b),
and $U=0.5$ and $U_1=1$ in (c) and (d). In these two examples,
$|J_1|\gg|J_{i\neq1}|$ outside the neighborhood where $J_1=0$. We
note that $G_1$ is considerably smaller than $G_0$ for any $\Delta$.

Let us comment on the elaborate structure of the system. For large
detunings, $|\delta_1|\gg\lambda$ and $|\Delta|\gg
\lambda,\,\Omega$, atoms in state $|1\rangle$ moves in an effective
potential $U_{eff}(x)\propto\cos^2(x)$. This is the common adiabatic
dispersive situation utilized in most experiments. The effective
potential in such case is obtained via adiabatic diagonalization of
the Hamiltonian. It is known that such approach gives rise to
non-adiabatic corrections, which can be expressed in terms of
effective gauge fields \cite{gauge}. However, in the dispersive
regime, the non-adiabatic corrections are vanishingly small and can
be neglected. On the other hand, in the intermediate regime, which
we are interested in, these corrections cannot be overlooked and
must be taken into account. In fact, only in the adiabatic limiting
situations $|\Delta|\rightarrow\pm\infty$ can an effective potential
be assigned to the internal states of the atoms
\cite{jonasjani,diabat}. Only in this limit will the boson operator
$\hat{b}_i^\dagger$ create an atom in a bare state
$\psi_{atom}(x-x_i)=w_1(x-x_i)|1\rangle$ or in a bare state
$\psi_{atom}(x-x_i)=w_2(x-x_i)|2\rangle$, otherwise both internal
states are populated. Three different energy contributions drive the
system, internal and kinetic atomic energies and atom-atom
interaction energy. This is not evident in the second quantized
formalism (\ref{eham}), containing only two parameters $J_1$ and
$G_0$. However, the internal atomic energy is indirectly embodied in
these two parameters, since the internal state greatly influences
the spinor Wannier functions.

\section{Mott-superfluid phases}\label{sec4}
In the previous section we found that the hopping coefficient may
change sign as $\Delta$ is varied. Thus, the effective hopping,
$J_1/G_0$, can be tuned from large negative to large positive
values. As will be described below, the change of sign of $J_1$ is
an outcome of a first order PT. Across this critical point, the
internal atomic state changes and thus defines two different phases.
In addition, in the superfluid phase, the sign of $J_1$ affects the
character of the many-body atomic state due to the phase matching
between the Wannier functions in the expansion (\ref{wanexp}).

For the phase diagrams, the sign change in $J_1$ implies that both
positive and negative regimes for the hopping should be considered,
and not restricted to only positive as in the regular Bose-Hubbard
model \cite{fisher}. We will present typical examples of the phase
diagrams in the $\mu-\Delta$ plane rather than in the $\mu-J_1$
plane, since experimentally $\Delta$ is an easily controllable
parameter. The phase diagrams are achieved by applying the {\it
strong coupling expansion} \cite{sce}, which has turned out to
reproduce accurate results for the Mott boundaries of the BH model
in one dimension.

\subsection{Phase diagrams}\label{ssec4b}
The parameters are derived from the numerically obtained spinor
Wannier functions. Moreover, $U$ and $U_1$ are chosen in such a way
that we can safely impose the above discussed approximations (see
also the following subsection).

\begin{figure}[ht]
\begin{center}
\includegraphics[width=8cm]{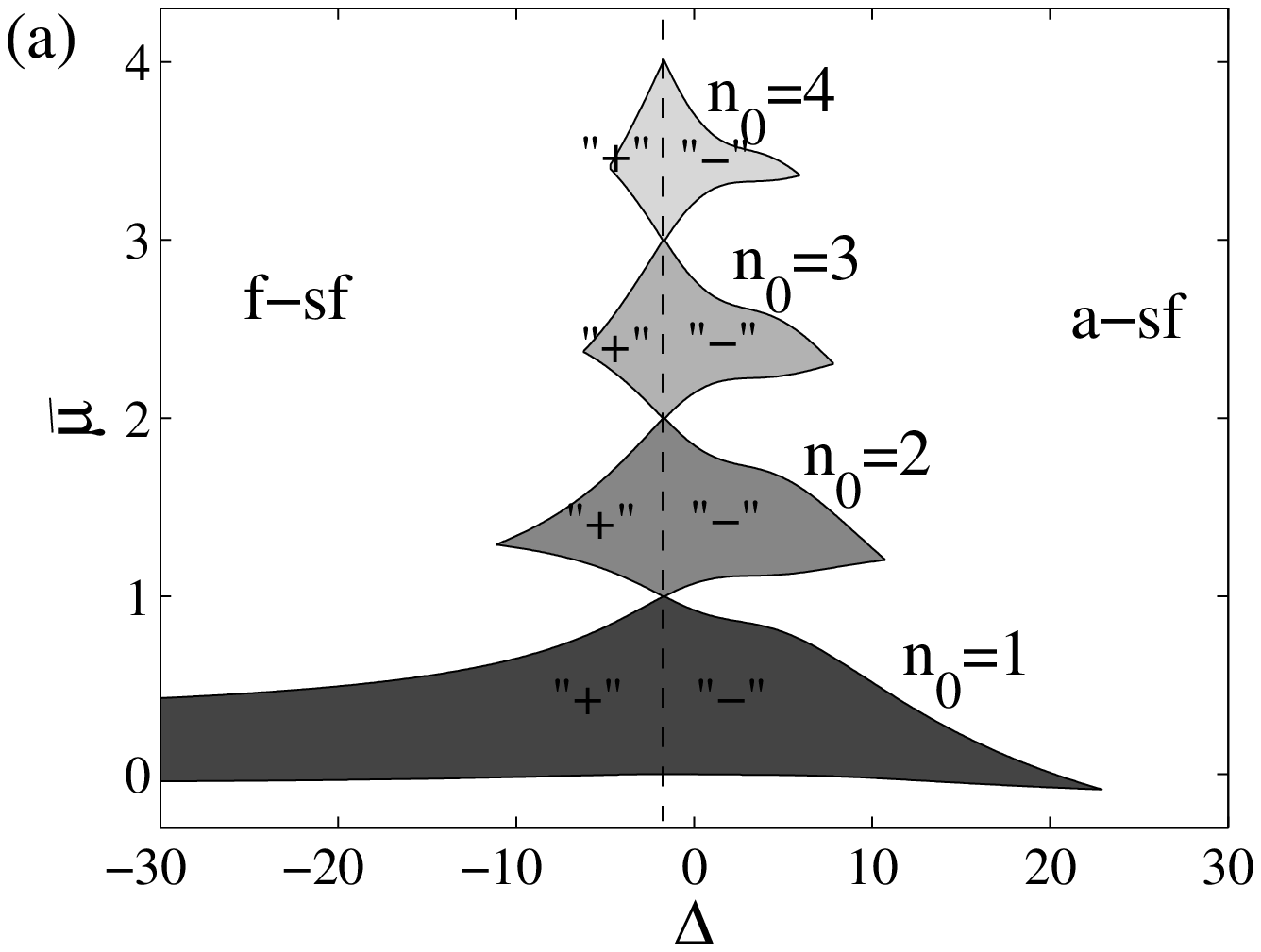}
\includegraphics[width=8cm]{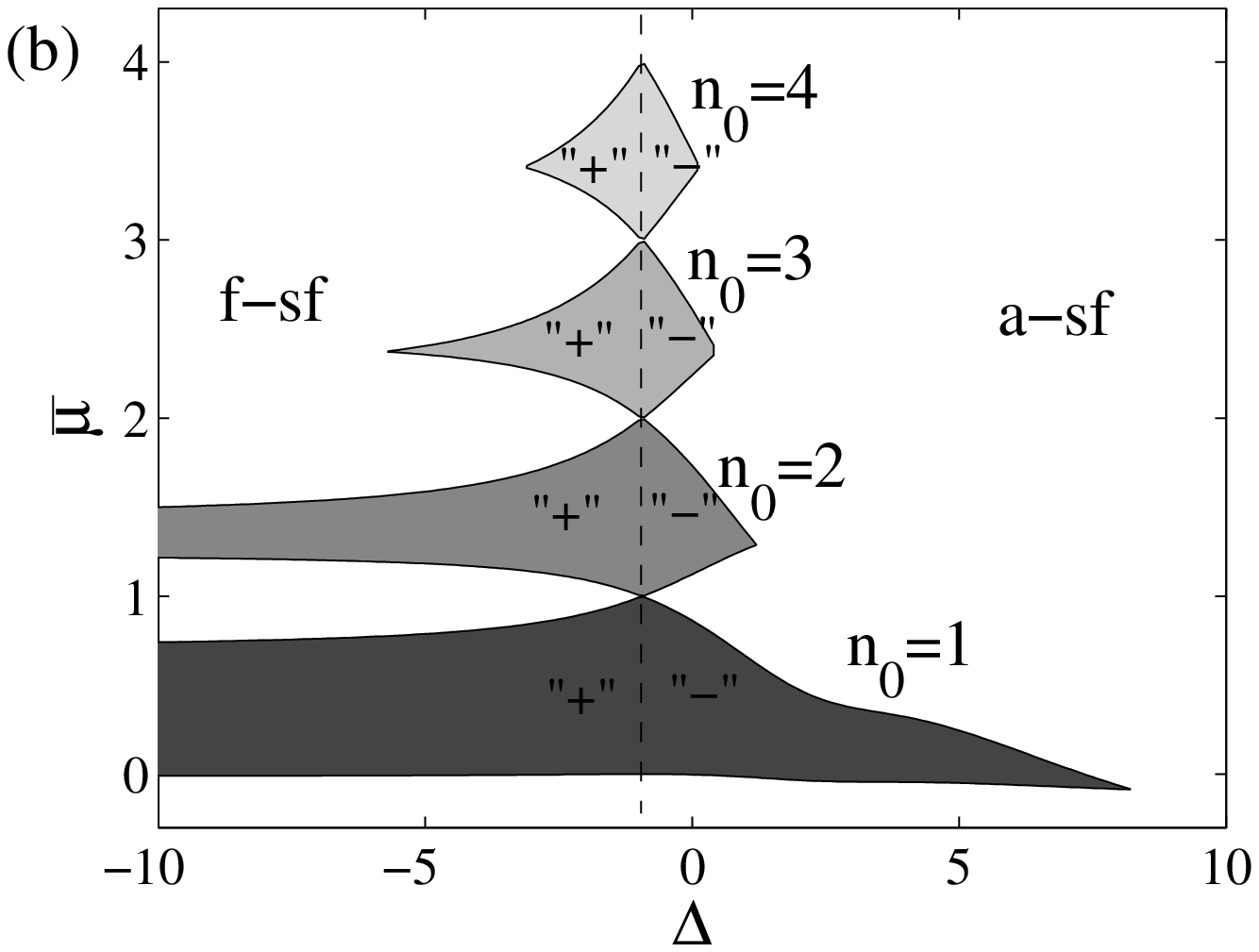}
\includegraphics[width=6cm]{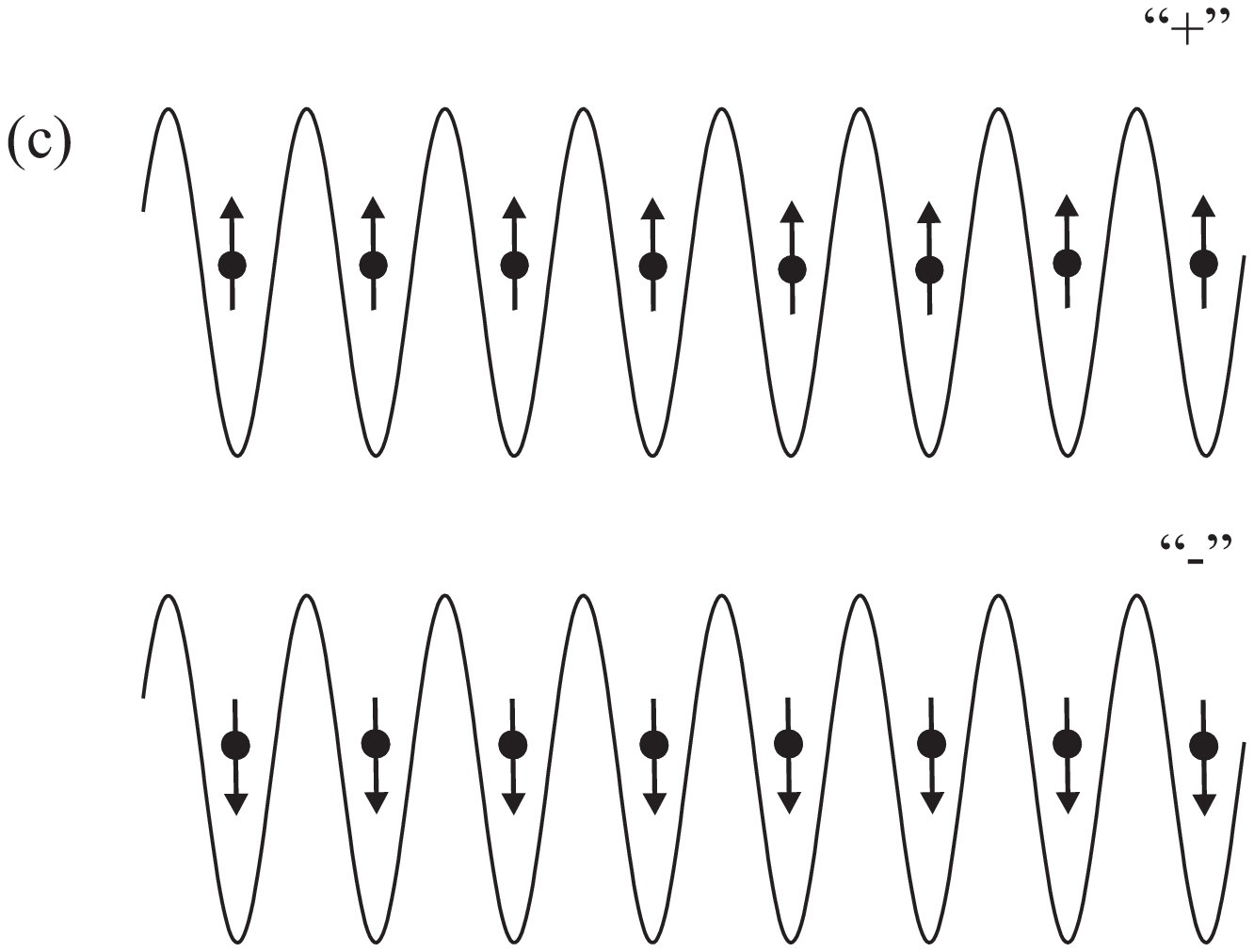}
\caption{Upper two plots (a) and (b) display the phase diagrams
(showing the first four Mott lobes) corresponding to Fig.~\ref{fig2}
(a) and (c). The vertical dashed lines is the phase boundary between
antiferromagnetic (a-sf) or ferromagnetic superfluid states (f-sf)
and between ``+'' and ``-'' Mott phases. The lower plot (c) gives a
schematic picture of the ``+'' and ``-'' Mott states by describing
the internal atomic states by their respective Bloch vectors.   }
\label{fig3}
\end{center}
\end{figure}

The phase diagrams obtained for the parameters of Fig.~\ref{fig2}
(a) and (c) are displayed in Fig.~\ref{fig3} (a) and (b)
respectively. The plot only shows the first four Mott lobes. The
vertical dashed line gives the crossover between positive and
negative $J_1$; left of the line $J_1>0$ and right of it $J_1<0$. We
note the asymmetry of the Mott zones on each side of the dashed
line. This irregularity originates from the Stark shift term
$U_1\cos(2x)$ appearing in the Hamiltonian (\ref{ham2}), which
effectively shift the resonance condition that defines the critical
detuning $\Delta_c$. Without this term \cite{foot}, the Mott lobes
are symmetric with respect to the dashed line.

In order to identify the different phases of Fig.~\ref{fig3} (a) and
(b), we first need to understand the underlying physics of the
single particle Hamiltonian (\ref{ham2}). In Ref.~\cite{jonasjani}
we demonstrated that a first order PT is obtained in an ideal gas of
coupled two-component bosonic or fermoinic atoms, when the detuning
is varied across a critical value. In the case of fermions, this
describes a topological PT, as the Fermi surface changes topology
across the critical point; each atom changes its momenta by either
$\pm1$. Changing the internal states of the atoms also shift the
atomic momenta by a multiple of the unit momentum. This yields a
competition between the two terms; for certain parameters it is more
favorable to lower the internal atomic energies, while in other
situations it is rather the atomic kinetic energy that should be
minimized. When crossing the critical point for this PT, which
occurs exactly when the hopping changes sign, each atom shifts their
momenta by $\pm1$ and their internal state populations are swapped.
Thus, the collective atomic inversion, which gives the population
imbalance between the internal atomic states, works as an order
parameter \cite{jonasjani}.

The atomic structure of the different phases become more clear by
thinking of the two-level atoms as spin 1/2 particles. The single
particle wave function is written as
\begin{equation}
\Phi(x)=\psi_1(x)|1\rangle+\psi_2(x)|2\rangle
\end{equation}
giving the Bloch vector components
\begin{equation}
\begin{array}{l}
u\equiv\langle\hat{\sigma}_x\rangle=2\mathrm{Re}[P_{12}],\\ \\
v\equiv\langle\hat{\sigma}_y\rangle=-2\mathrm{Im}[P_{12}],\\ \\
w\equiv\langle\hat{\sigma}_x\rangle=P_{11}-P_{22},
\end{array}
\end{equation}
where $P_{ij}=\int\psi_i^*(x)\psi_j(x)dx$. The last component of the
Bloch vector, $w$, is simply the atomic inversion. From the
decoupling of the two basis sets (\ref{basis}), it follows that
whenever only one bases set is populated $P_{ij}=0$ unless $i=j$.
Thereby, $u=v=0$ in the ``+'' Mott and the ``-'' Mott states, and
the corresponding atomic Bloch vector points either towards the
north or the south pole of the Bloch sphere respectively, as
schematically shown in Fig.~\ref{fig3} (c). The fact that the length
of the vector is in general smaller than unity reflects the
entanglement shared between internal and motional atomic degrees of
freedom.

The same type of PT exists also for incommensurate fillings
(superfluid states). However, contrary to atoms in the Mott state,
the superfluid states possess as well an inherent coherence within
the atoms and one finds that the PT manifests itself in this
coherence. In particular, such transition corresponds to going from
an antiferromagnetic (a-sf) to a ferromagnetic superfluid state
(f-sf), or vice versa. The terms antiferro- and ferromagnetic come
from phase matching between site wave functions in the expansion
(\ref{wanexp}). For the antiferromagnetic state, there is a
$\pi$-phase sign-flip between neighboring Wannier functions. We term
these superfluid states a-sf or f-sf in Fig.~\ref{fig3}. Note that
the terminology of antiferro and ferromagnetic states originates
from the phase-matching between Wannier functions, and not from the
spin orientation (Bloch vectors) between neighboring sites.

The analysis has been carried out by restricting the atomic states
to the set $\{|\varphi_\eta(q)\rangle\}$ of Eq.~(\ref{basis}). For
the many-body system, one must take atom-atom scattering into
account. These contributions break the symmetry defined by the
operator $\hat{I}$ given in Eq.~(\ref{sym}), and consequently cause
coupling between the two sets of (\ref{basis}). The approach
utilized in this work does not include such population transfer
processes. We will now argue that these are indeed very small even
at very large effective scattering amplitudes $g$ (and $g_{12}$). To
do so we solve the corresponding Gross-Pitaevskii equation
(\ref{gp}) obtained by replacing the atomic operators by mean-field
wave functions. We start from an ansatz wave function for $\Psi(x)$
which completely resides in one of the bases sets
$\{|\varphi_\eta(q)\rangle\}$ or $\{|\phi_\eta(q)\rangle\}$. Given
any set of parameters $U,\,U_1,\,\Delta,$ and $\mathbf{g}$, the
ground state is obtained via imaginary time-evolution. If
$\mathbf{g}=0$, no population transfer occurs between the two bases
sets and the obtained ground state populates only one basis set. For
the field amplitudes $U$ and $U_1$ of Fig.~\ref{fig3} and various
$\Delta$, effective couplings $g=100$ and $g_{12}=0$ give a
population imbalance between the bases sets for the ground state of
$<5/1000$. That is, even for effective scattering amplitudes as
large as 100 recoil energies, more than 99.5 $\%$ of the ground
state population resides in one of the basis sets. It turns out that
the amount of coupling between the two bases sets is more sensitive
to the strength of $g_{12}$ than to the one of $g$. For
$g_{12}=100$, about 10 $\%$ may populate the orthogonal basis set
(population imbalance 1/9). On the other hand, we find that $w$ does
not depend on $g_{12}$, and therefore a non-zero $g_{12}$ leaves the
order parameter $w$ unchanged. We may conclude that in terms of the
Bloch vector, scattering between the atoms induces only a small
deviation of the vector from the $z$-axis, but the projection onto
the $z$-axis is very insensitive to these scattering processes. We
should point out that solving the Gross-Pitaevskii equation is only
justified deep in the superfluid regime, but nonetheless, we believe
that the above analysis utilizing very large scattering amplitudes
demonstrates the robustness of the assumption of neglecting coupling
between the two bases sets (see also \cite{anssi}).

In Ref.~\cite{jonasjani} we showed that the anomalous structure of
the dispersions is also encountered in a two dimensional lattice.
Furthermore, in \cite{anssi} we studied the dynamics of the
mean-field PT deriving from the atypical dispersions in both one and
two dimensions, and proved its persistence in two dimensions.
Thereby, it is clear that our findings are not restricted to the
special case of a one dimensional lattice, but apply also to higher
dimensions. However, in the present work, studying higher dimensions
would be considerably more cumbersome as the Hamiltonian cannot be
written as a simple sum of lattice potentials in the different
spatial directions, and consequently calculation of the
corresponding Bloch and Wannier functions are not as straight
forward as for a regular dispersive square or cubic lattice.
Moreover, the strong coupling expansion used for determining the
phase diagrams in the proceeding section only provides
quantitatively accurate results in one dimension, while in two and
three dimensions it reproduces only qualitative results.

\subsection{Validity of approximations}\label{ssec4c}
Various approximations have been imposed in order to derive the
phase diagrams; tight binding and single band approximations, and
truncating the strong coupling expansion at third order. In this
subsection we systematically discuss the justification of such
assumptions. A rule of thumb is that these approximations are all
related in the regular Bose-Hubbard model and share more or less the
same validity regimes \cite{jonasbcn}. It is not deducible that this
holds also for our two-component model where for instance the
Wannier functions are spinors.

Already Fig.~\ref{fig1} justify the application of the tight binding
approximation. This, of course, is due to our choice of parameters.
As a second check of the tight binding approximation, we have
modified the strong coupling expansion to include tunneling
processes beyond nearest neighbors and recalculated the phase
diagrams of Fig.~\ref{fig3} and found only minimal corrections. Note
that when $J_1=0$, the tight binding approximation in general fails.
In this regime, however, the hopping terms beyond nearest neighbor
are, for the examples presented in this paper, very small such that
the dynamics is predominantly driven by the onsite interaction and
therefore the system must be in a Mott state.

The justification for the use of the single band approximation has
been investigated by evaluating overlap integrals between spinor
Wannier functions of the first and the second band. For
Fig.~\ref{fig3}, we found that these coupling elements are
everywhere much smaller than $J_1$ and $G_0$ except for
$\Delta\approx0$ in (b). Thus, tuning $\Delta$ non-adiabatically
across resonance, when $U=0.5$ and $U_1=1$, may cause population of
excited bands.

Finally, in order to discuss on what grounds the third order strong
coupling expansion may be applied, we have compared the phase
diagrams of Fig.~\ref{fig3} with the ones obtained by utilizing
second order perturbation theory instead of third order and found
only slight modifications.

\section{Character of different phases}\label{sec5}
As pointed out in the introduction, coupled two-level atoms in
optical lattices were first studied in Ref.~\cite{multicomp1},
considering two optical lattices driving a Raman transition in
$\Lambda$ atoms. By tuning the relative phase between the two
lattices, it was demonstrated that the hopping coefficient in the
corresponding Bose-Hubbard Hamiltonian could attain negative as well
as positive values. Furthermore, due to the phase factor in the
Wannier expansion (\ref{wanexp}), it follows that the coherence in a
superfluid state is different depending on the sign of the hopping
coefficient. In particular, if the phase difference between
consecutive Wannier functions in (\ref{wanexp}) is $\pi$ the
superfluid state was termed antiferromagnetic, while a zero phase
difference characterizes ferromagnetic superfluid states. In
Ref.~\cite{multicomp1}, it was also predicted that ballistic
expansion of an antiferromagnetic or ferromagnetic superfluid state
would render dissimilar time-of-flight measurements. Although, in
principle, measuring the atomic population inversion will determine
the character of the superfluid states, in this section we analyze
the corresponding time-of-flight scenario in our model. In other
words, we investigate if the different phase modulation of the
superfluid states is sufficient for separating between the two
possible states.

\begin{figure}[ht]
\begin{center}
\includegraphics[width=8cm]{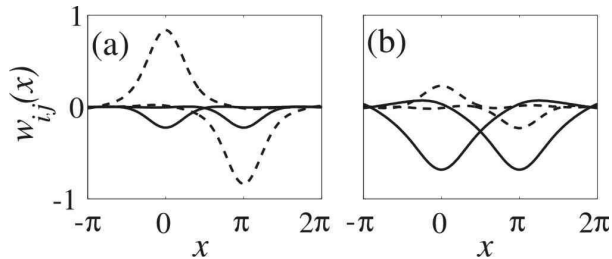}
\caption{Neighboring constituent Wannier functions $w_{j,1}(x)$
(dotted) and $w_{j,2}(x)$ (solid). In (a) $J_1\approx0.1$
corresponding to $U=1$, $U_1=0.5$, $\Delta=-11.5$, while
$J_1\approx-0.1$ in (b) obtained by choosing $U=1$, $U_1=0.5$,
$\Delta=11.5$. Note that the dashed but not the solid line flip
phase between neighboring sites. All parameters are dimensionless.} \label{fig4}
\end{center}
\end{figure}

\subsection{Effect of positive and negative hopping coefficients}\label{sec5a}
In the internal $\left\{|1\rangle,\,|2\rangle\right\}$ basis, the
spinor Wannier function at site $i$ is decomposed as
\begin{equation}\label{wanspin}
{\bf w}_j(x)=\left[\begin{array}{c}w_{j,1}(x)\\w_{j,2}(x)\end{array}\right].
\end{equation}
Fig.~\ref{fig4} visualizes two examples of the neighboring Wannier
functions. In (a), the nearest neighbor tunneling coefficient is
positive, $J_1\approx0.1$, while in (b) it is negative but with the
same amplitude, $J_1\approx-0.1$. The figure evidences the swapping
of internal state populations between positive and negative hopping.
Consequently, an internal state selective measurement distinguishes
between antiferromagnetic or ferromagnetic states and between Mott
``+'' and Mott ``-'' states. Nonetheless, in the next subsection, we
also show how a time-of-flight detection can tell apart the two
different superfluid states.

\subsection{Time-of-flight detection}\label{sec5b}

\begin{figure}[ht]
\begin{center}
\includegraphics[width=8cm]{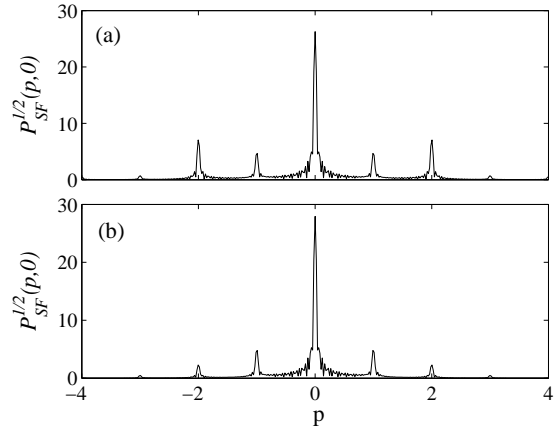}
\caption{The momentum distributions of the state (\ref{atstate})
with $\Delta=-11.5$ (a) and $\Delta=11.5$ (b). The other
dimensionless parameters are as in Fig.~\ref{fig4}. The peaks at
$p=\pm2$ are more pronounced in (a) than in (b) giving different
interference patterns of the freely expanding atomic condensates. }
\label{fig6}
\end{center}
\end{figure}

Our idea is to study the different interferences induced by having 0
or $\pi$ phase correlation between neighboring sites. Thus, we have
a matter wave function as
\begin{equation}\label{atstate}
\Psi_{SF}(x,0)=\sum_je^{i\theta_j}\left[\begin{array}{c}w_{j1}(x)\\w_{j2}(x)\end{array}\right]
\end{equation}
and study the impact of having
\begin{equation}
\theta_{j}=\left\{\begin{array}{ll}
0,\hspace{0.5cm} & J_1>0 \\
j\pi,\hspace{0.5cm} & J_1<0
\end{array}\right.
\end{equation}
when the wave function is freely evolving. The ballistically
expanded wave function reads
\begin{equation}
\Psi_{SF}(x,t_{tof})=e^{-i\hat{p}^2t_{tof}}\Psi_{SF}(x),
\end{equation}
where $t_{tof}$ is the time-of-flight time between release of the
superfluid state till measurement of it. The total probability
distribution
\begin{equation}\label{totdist}
P_{tot}(x,t_{tof})=|\Psi_{SF}(x,t_{tof})|^2
\end{equation}
or its constituent probability distributions $P_i(x,t_{tof})$
($i=1,\,2$) for the atomic internal states are assumed being detected
after the ballistic expansion. For $t_{tof}\rightarrow\infty$, the
momentum distribution $P_{SF}(p,t=0)$ corresponding to
$\Psi_{SF}(x,0)$ in Eq.~(\ref{atstate}), is encoded into the
distribution $\Psi_{SF}(x,t_{tof}=\infty)$. In Fig.~\ref{fig6}, we
display the square root of the momentum distributions of
$\Psi_{SF}(x,0)$. In (a) we use the parameters of Fig.~\ref{fig4}
(a) (and thus a constant phase between the Wannier functions), while
in (b) the parameters are as in Fig.~\ref{fig4} (b) ($\pi$ phase
modulation between the neighboring Wannier functions). For $J_1>0$,
the even numbers of momenta are more strongly populated; the
peaks around $\pm2$ are more distinct in Fig.~\ref{fig6} (a) than in
(b). Note that for this example $J_1$ have the same strength in both
cases, but the corresponding Wannier functions (see Fig.~\ref{fig4})
are different. Thus, it is not only the phase $\theta_j$ which
distinguishes the two cases.

\begin{figure}[ht]
\begin{center}
\includegraphics[width=8cm]{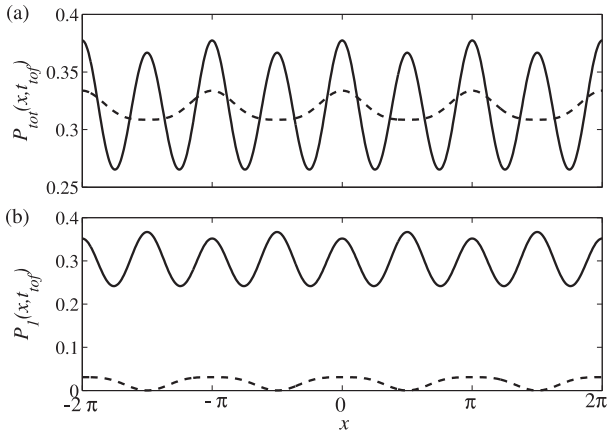}
\caption{Atomic distributions $P_{tot}(x,t_{tof})$ (a) and
$P_1(x,t_{tof})$ (b) after a time-of-flight spreading $t_{tof}=4$.
For solid lines; $\Delta=11.5$, while for dotted lines;
$\Delta=-11.5$, and in both cases $U=1$ and $U_1=0.5$. This set of
parameters give $J_1=0.1$ and $J_1=-0.1$ respectively. The
difference between solid and dotted lines derives from the different
Wannier functions of the two cases, but also from the phase factor
$e^{i\theta_j}$ in the Wannier expansion (\ref{wanexp}).  }
\label{fig7}
\end{center}
\end{figure}

The difference in momentum distributions will also manifest itself
in the position distributions $P_{tot}(x,t_{tof})$ and
$P_i(x,t_{tof})$ for finite times $t_{tof}$. The results for the
total probability distribution (\ref{totdist}) and the distribution
for the internal state 1,  $P_1(x,t_{tof})$, are depicted in
Fig.~\ref{fig7} (a) and (b) respectively. The time-of-flight
$t_{tof}=4$, guaranteeing that the interference has been well
established. Noticeable from the figure is that for positive hopping
the distribution shows a super-structure with two local maxima for
each period, not seen for $J_1<0$. The great difference in
probability amplitude between the internal atomic states in (b)
derives from the fact that the two internal states are unequally
populated due to the different detunings; $\Delta=11.5$ and
$\Delta=-11.5$.

\section{Conclusions}
In this work we have presented an analysis of a gas of coupled
two-level bosonic atoms in an one dimensional optical lattice. The
spectrum of the single particle Hamiltonian was found to possess
peculiar characteristics originating from the coupled dynamics. In
an earlier work, we demonstrated that PTs can be obtained both for
fermionic and bosonic atoms in the current model in the absence of
atom-atom interaction \cite{jonasjani}. Including scattering between
the atoms, as in this paper, we identified the PT of
Ref.~\cite{jonasjani} as a sign change in the site hopping
parameter. The corresponding PT was shown to be between
distinguishable Mott or superfluid states, characterized by an
imbalance of the population of internal atomic states. Moreover, a
thorough analysis about the effect of a positive or a negative
nearest neighbor tunneling coefficient was given, focusing on
time-of-flight detection of the condensate. The model can be
generalized to two and three dimensions, and due to the anomalous
dispersions found also in higher dimensions~\cite{jonasjani,anssi},
the corresponding phase diagrams is believed to show similar
structures. However, in higher dimensions the strong coupling
expansion method utilized in this work is supposed to give less
reliable results, and other approaches would then be preferable.

The imposed approximations and their validity regimes were studied.
The present paper restrict the analysis to regimes where these
approximations are justified. However, it is expected that new
phenomena will occur beyond such limitations, not encountered in the
regular Bose-Hubbard Hamiltonian. For example, the nearest neighbor
hopping may vanish, and consequently long range or semi-long range
interaction might become important. We are currently investigating
regimes outside single band and tight binding approximations by
using different methods. These results are left for future
publications. We are also studying the dynamics of the condensate as
the system is driven through the critical point $J_1=0$
\cite{anssi}.

\begin{acknowledgements}
JL acknowledge support from the MEC program (FIS2005-04627).
\end{acknowledgements}


\begin{thebibliography}{999}
\bibitem{diabat} Defining a diabatic representation and diabatic limit in the present model is non-trivial.
\bibitem{foot} Discarding this term from the Hamiltonian gives a model in which the two atomic states are directly coupled by the optical lattice, see \cite{jonasjani}.
\end{thebibliography}
\end{document}